\documentclass[aps,nobibnotes,preprint,groupedaddress,showkeys,showpacs]{revtex4}
\usepackage{graphicx}
\usepackage{dcolumn}
\usepackage{amsmath}
\usepackage{amssymb}
\usepackage{bm}

\begin{document}
\input{epsf}

\preprint{APS/123-QED}

%
%
%
%

\title{State-resolved rotational cross sections and thermal rate coefficients for
$ortho$-/$para$-H$_2$+HD at low temperatures and HD+HD elastic scattering}

\author{Renat A. Sultanov}
\email{rasultanov@stcloudstate.edu, r.sultanov2@yahoo.com}
\affiliation{Business Computing Research Laboratory, St. Cloud State
University, 367B Centennial Hall, 720-4th Avenue South, St Cloud,
MN 56301-4498, USA}

\author{Avas V. Khugaev}  
\email{khugaev@inp.uz, avas@iucaa.ernet.in}
\affiliation{
Institute of Nuclear Physics, Uzbekistan Academy of Sciences\\ 100214
Tashkent, Uzbekistan}  

\author{Dennis Guster}  
\email{dcguster@stcloudstate.edu}
\affiliation{Business Computers and Information Systems, St. Cloud State
University, 367C, Centennial Hall, 720-4th Avenue South, St Cloud,
MN 56301-4498, USA}

\date{\today}

\begin{abstract}
\noindent Results for quantum mechanical calculations of the integral cross
sections and 
corresponding thermal rate coefficients for para-/ortho-H2+HD
collisions are presented. 
Because of significant astrophysical interest in regard to the cooling 
of primodial gas the low temperature limit of para-/ortho-H2+HD is
investigated. 
Sharp resonances in the rotational state-resolved cross sections have 
been calculated at low energies. These resonances are important and 
significantly contribute to the corresponding rotational
state-resolved 
thermal rate coefficients, particularly at low temperatures, that is
less than $T \sim 100$K. 
Additionally in this work, the cross sections for the elastic HD+HD
collision have also been calculated. 
We obtained quite satisfactory agreement with the results of other theoretical works and experiments.
\end{abstract}



\maketitle
\section{Introduction}
\label{intro}
We report rotational state-resolved cross sections
and corresponding thermal rate coefficients for the fundamental elastic and
inelastic collisions between hydrogen molecules in both $ortho$- and $para$- states
and its isotopic partner, the HD molecule:
\begin{equation}
ortho\mbox{-}/para\mbox{-H}_2(j_2) +\mbox{HD}(j_1) \rightarrow ortho\mbox{-}/para\mbox{-H}_2(j'_2) +
\mbox{HD}(j'_1).
\label{eq:h2hd}
\end{equation}
Here H is hydrogen and D is deuterium. 
HD is the second most abundant primordial molecule after H$_2$
and plays an important role in the chemistry of the interstellar medium \cite{dalgar72},
more specifically in rotational energy transfer collisions (\ref{eq:h2hd})
\cite{schaefer90,flower99a,renat07}.
Also, HD itself is of interest  
as a light molecule with a small but constant dipole moment $|\vec d| =
(8.1\pm 0.5)\cdot 10^{-4}$ D \cite{wishnow92}.
This ability of HD makes the study of this molecule an important
undertaking in astrophysics and other fields, such as in
condensed matter physics, see for instance \cite{strzh06}.

Furthermore, H$_2$ and HD molecules play a significant role   
in the formation of the primary stars in the Universe \cite{volker02,volker04}, where
the dynamics of H$_2$+ H$_2$, H$_2$+ HD and HD+HD molecular collisions
can make a substantial contribution to the formation of the primordial
structure of stars \cite{greer08}.
The spectroscopic measurements from distant quasars
with H$_2$ and HD molecules in the cold gases around them or on the
line of sight can be very important and requires detailed knowledge of the cross sections
at low energies \cite{ge97,varshalovich01}.

Although the field of molecular collisions between hydrogen molecules and its different
isotopic combinations: H$_2$+D$_2$, D$_2$+HD, HD+HD and others,
have already a rather long research history
\cite{gentry77,johnson79,gelb79,billing92,buck83,chandler86,chandler88},
there are substantial discrepancies between the results of different
research groups. The recent appearance of the new potential energy surface for the
H$_4$ system makes it very useful to carry out new
calculations for these important systems.

By now the deuterium chemistry of the early Universe has been
extensively studied by a few research groups, see for example,
\cite{schaefer90} and \cite{flower99a,renat07}.
In these works the authors carried out quantum-mechanical
state-resolved calculations  for a wide range of temperatures.
Schaefer's work \cite{schaefer90} presents results for the thermal
rate coefficients at
very low temperatures down to $\sim$10 K. The corresponding
cross sections have been calculated at collisional velocities $\sim 12$
m/s, which are almost equal to ultra-low energies.

The author of work \cite{flower99a} used Schwenke's modified H$_2$-H$_2$ 
potential \cite{schwenke90,schwenke88}. Schaefer applied a different
surface, which was also a modification of older PESs, see for example \cite{schaefer87}.
However, as we already mentioned, the
importance of the H$_2$+HD rotational energy transfer 
collisions makes it very reasonable to carry out new calculations for
the systems with the use of the BMKP PES published in 2002
which provides the latest and presumably most accurate
H$_4$ surface \cite{booth02}.


In this work the scattering cross sections and their corresponding thermal rate coefficients 
are calculated using a non reactive quantum-mechanical close-coupling
approach.  In the work \cite{renat07} we already presented the details
of the method we used, therefore in this work we continue beyond that point
with the results for $ortho$-H$_2$+HD.
The calculations are done for very low and low temperatures from 5 K
to up to 300 K. Atomic units $(e=m_{e}=\hbar=1)$ are used throughout these calculations.
\section{Results}  
\label{Results}
Before undertaking production calculations we carried out a
large number of test calculations to insure the convergence of the results with respect to all 
parameters that enter into the propagation of the Schr\"odinger
equation \cite{hutson94}. The details of the quantum-mechanical
computation approach are presented in \cite{green75}. The same test calculations were
carried out in our previous works \cite{renat06a,renat06b} for 
the $ortho$-/$para$-H$_2$+H$_2$ collisions,
which involved the intermolecular distance $R_3$, the total angular
momentum $J$ of the four atomic system, 
the number of rotational levels to be included in the close coupling 
expansion $N_{lvl}$ and others.

In these calculations the {\it VRTP} mechanism has been used, which allows us to specify 
$V(\vec R_1,\vec R_2,\vec R_3)$ explicitly rather
than to expand the potential in angular functions, see the 
MOLSCAT manual \cite{hutson94}. As in our previous works
we reached convergence for the integral cross sections,
$\sigma(j'_1,j'_2;j_1j_2,\epsilon)$, in all considered
collisions. However, we have found in this work, that in the case
of $ortho$-H$_2$+HD the convergence is slower than in the case of $para$-H$_2$+HD.


\subsection{$ortho$-/$para$-H$_2$+HD and HD+HD cross sections}
\label{orthoH2HD}

In Fig. 1 we present the results for the low energy
$ortho$-/$para$-H$_2$ + HD and HD+HD elastic scattering collisions. In these calculations
the collision energy ranges from 10$^{-5}$ cm$^{-1}$ to 100 cm$^{-1}$. As can be seen at energies less than 
$\sim 5 \cdot 10^{-4}$ cm$^{-1}$ the elastic cross sections, $\sigma_{el}$,
display almost constant values.
This result is in good agreement with the non-relativistic quantum-mechanical
scattering theory, that is at very low collision energies:
\begin{equation}
\sigma_{el}(E_{coll}\sim 0) = \pi a_0^2 =const,
\end{equation}
where $a_0$ is the scattering length \cite{landau}. From our calculations
in the case of $ortho$-H$_2$+HD: $\sigma_{el}=\pi a_0^2=380.70\cdot 10^{-16}$ cm$^2$.
The corresponding scattering length 
is
$a_{0}(ortho\mbox{-H}_2+HD) = 11.01\cdot 10^{-8}$ cm.
A rather sharp resonance is found in this cross section at energies
around 0.56 cm$^{-1}$.  The value of this resonance is 
$\sigma_{res}(ortho\mbox{-H}_2+HD) = 1,897.8\cdot 10^{-16}$ cm$^2$.

Below, in the same figure we also present the results from a
$para$-H$_2$+HD scattering cross section. These results were not shown in \cite{renat07}.
Further, we computed the HD+HD elastic collision.
These calculations are also done with the BMKP PES.
Fig.\ 1 shows this result. It is seen, that the behaviour of
all three cross sections is almost identical, although the values are rather different.
Again, the collision energy ranges from 10$^{-5}$
cm$^{-1}$ to 100 cm$^{-1}$. At energies less than 
$\sim 5 \cdot 10^{-4}$ cm$^{-1}$ the elastic cross sections
display almost constant values:
$\sigma_{el}(para\mbox{-H}_2+HD)=380.031\cdot 10^{-16}$ cm$^2$, and
$\sigma_{el}(\mbox{HD+HD})=274.84\cdot 10^{-16}$ cm$^2$.
The corresponding scattering lengths are:
$a_{0}(para\mbox{-H2+HD})=11.0\cdot 10^{-8}$ cm and
$a_{0}(\mbox{HD+HD})=9.35\cdot 10^{-8}$ cm.

Sharp resonances have been calculated at energies
around 0.57 cm$^{-1}$ in the case of $para$-H2+HD
and around 2.05 cm$^{-1}$ for HD+HD.  The values of these resonances are
the following: 
$\sigma_{res}(para\mbox{-H}_2+HD)=1,888.84\cdot 10^{-16}$ cm$^2$
and $\sigma_{res}(\mbox{HD+HD})=737.43\cdot 10^{-16}$ cm$^2$.

Next, one can see from Fig.\ 1, that the elastic cross sections are
closely related to each other and the calculated values of $a_0$'s
have reasonably similar values. However there are significant differences
in the positions and the values of the resonances
between $ortho$-/$para$-H$_2$+HD and HD+HD collisions.
The reason for these differences is due to the permanent dipole
moment of the HD molecules. This phenomenon is important at low
energies and is responsible for the energy shift
in the total Hamiltonian of the four-atomic systems. The additional interaction
makes the shift of the position of the high value resonance in
HD+HD elastic cross section
relative to the positions of the resonances in the $ortho$-/$para$-H$_2$+HD systems.


In addition to the results of Fig.\ 1 in Fig.\ 2 our HD+HD elastic cross sections are shown
together with the corresponding results from    
paper \cite{johnson79}. Please see Fig.\ 16 and Fig.\ 17 of \cite{johnson79}, where theoretical
and some experimental results for HD+HD are presented.
It is seen, that the trend of the both cross sections
is similar to each other, this is important, although the values are rather
different at some collisional energies (velocities).
We consider these results as a quite sufficient test of our calculations for the
HD+HD collisions.

Fig.\ 3 shows results for the state-resolved cross sections $\sigma(j'_{1}j'_{2};j_{1}j_{2})
(\epsilon)$ in the $ortho$-case of the collision (1).
In the upper and middle plots we present our data for the
$(j_1=1,j_2=1)\rightarrow (j'_1=0,j'_2=1)$ 
and $(j_1=2,j_2=1)\rightarrow (j'_1=j'_2=1)$ 
quantum transitions together with the results of Schaefer's work
\cite{schaefer90},
which applied a modified version of Schwenke's potential 
\cite{schwenke88} and a different dynamical quantum-mechanical method.
In the lower plot results for 
$(j_1=2,j_2=1)\rightarrow (j'_1=0,j'_2=1)$ are presented together
with the results of work \cite{schaefer90}. The values
of this channel cross section are lower by factor of 10 than in the
case of the two previous cross sections.

As can be seen for the transition quantum states
$(j_1=1,j_2=1)\rightarrow (j'_1=0,j'_2=1)$ 
and $(j_1=2,j_2=1)\rightarrow (j'_1=j'_2=1)$ the agreement between our
results and Schaefer's calculations is fairly good. However, for the 
$(j_1=2,j_2=1)\rightarrow (j'_1=0,j'_2=1)$ channel cross section we
found significant disagreements in the cross section. The difference
between our cross section and Schaefer's results reaches the factor of 2
at a collisional velocity around 270 m/s.

In conclusion, in the case of $ortho$-/$para$-H$_2$+HD
we reproduced all sharp resonances in the low energy region \cite{renat07}. These high
value cross sections should contribute significantly
to the cooling process of the astrophysical media, therefore it is of
crucial importance to carry out calculations for these cross sections accurately.
However, in the case of higher transition states we found some disagreements (up to 100\%) in the
cross sections with work \cite{schaefer90}. But these cross sections
have values of relatively low magnitude.


\subsection{$ortho$-/$para$-H$_2$+HD thermal rate coefficients}
\label{paraH2HDandHDHD}

In this section we present new results for thermal rate coefficients
regarding the case of $ortho$-/$para$-H$_2$+HD.
Corresponding cross sections have been presented in the previous
section and some of them have been published in work \cite{renat07}.
In \cite{renat07} we were able to reproduce all relevant resonances in the cross
sections for different rotational channels \cite{schaefer90}, however 
significant differences and new resonances have been found in the cross sections
with higher values of the rotational transition states.  

Fig.\ 4 shows temperature dependent thermal rate coefficients
corresponding to two relevant cross sections from Fig. 2: for
$(j_1=1,j_2=1)\rightarrow(j'_1=0,j'_2=1)$ and $(j_1=2,j_2=1)\rightarrow(j'_1=j'_2=1)$ 
quantum-mechanical rotational transition states. Again agreement with work
\cite{schaefer90} is very good. However there is significant
disagreement with results of work \cite{flower99a}.

Fig.\ 5 presents our results together with Schaefer's data from
\cite{schaefer90} for the
$(j_1=1,j_2=0)\rightarrow(j'_1=j'_2=0)$, $(j_1=2,j_2=0)\rightarrow(j'_1=1,j'_2=0)$
and $(j_1=2,j_2=0)\rightarrow(j'_1=j'_2=0)$ thermal rate coefficients.
One can see, that the first two thermal rates agree very well with the
results of work \cite{schaefer90}, however the third (lower plot) is
not in good agreement. As in Fig.\ 3, there is significant
disagreement with the corresponding results of work \cite{flower99a}.

In \cite{renat07} and in this work we provide 
cross sections and corresponding thermal rates for only relevant transition
states, that is transitions with relatively high cross sections. 
Detailed calculations for many other transition quantum numbers and,
probably, for collisions of other hydrogen isotopes should be done in
the future.

Through this analysis we can conclude, that the BMKP PES can reproduce the general behaviour 
of all considered cross sections and thermal rates in the $ortho/para$-H$_2$+HD
collisions. For the lower quantum states we obtained
sufficient agreement with the results of work \cite{schaefer90} 
and significant disagreement with \cite{flower99a}.
Also, for transition states with higher values the BMKP PES provides 
rather small cross sections and thermal rate coefficients
relative to the corresponding results of work \cite{schaefer90}.

\section{Conclusion}
\label{Conclusion}

In this work the state-resolved close-coupling quantum-mechanical calculations for rotational
excitation and de-excitation cross sections of the
$ortho$-/$para$-H$_2$+HD and HD+HD collisions are presented.
The latest, global BMKP surface for the H$_2$-H$_2$ system has been appropriately adopted for
H$_2$+HD and HD+HD. The linear rigid rotor model for the H$_2$ and HD molecules is applied.
A test of convergence and the results for cross sections with the BMKP PES
are obtained for a wide range of kinetic velocities including very low
values down to $\sim$10 m/s.
As in work \cite{renat07} these results revealed, that the applied
quantum-mechanical method together with the BMKP PES is able to
provide reliable cross sections and thermal rate coefficients in collisions (1).

An interesting result of this work lies in the comparison between our
calculations with the BMKP PES, calculations of \cite{flower99a}, and
with the results of \cite{schaefer90}. As we mentioned, these
works used pure quantum mechanical methods, but within the rigid rotor
approximation: the distances between hydrogen atoms in the
H$_2$ or HD molecules have been fixed. These works used different H$_4$
potential energy surfaces.
In Fig.\ 3 we showed that our results for the state-resolved cross
sections $\sigma_{nl \rightarrow n'l'}(E_{coll})$ are 
close to the corresponding cross sections of the relatively old work
\cite{schaefer90}. Furthermore, because the state-resolved thermal rate
coefficients $k_{nl \rightarrow n'l'}(T)$ are less sensitive to
different collision parameters, for instance, interaction potentials, 
for $k_{nl \rightarrow n'l'}(T)$ we obtained even better agreement
with the results of work \cite{schaefer90}. This is seen in Figs. 4 and 5.

However, the same type comparisons with the corresponding
thermal rate coefficients from the newer work \cite{flower99a} demonstrated
substantial disagreements, which are also seen in Figs.\ 4 and 5.
The deflection between results of \cite{flower99a} and
our data and Schaefer's results can reach almost 60\%.
In future works it will be very useful for comparison purposes to adopt other
precise H$_4$ PES from the work \cite{djpes00}  
and apply it to the ortho-/para-H$_2$+HD collisions. 
The Deep and Johnson potential energy surface (DJ PES) was already
successfully applied to H$_2$+H$_2$ 
in \cite{renat06a,renat06b,otto08}.     
Also, it seems to us that it would be very useful
to carry out new calculations for various differential cross sections 
$d \sigma_{nl \rightarrow n'l'}(\theta,\varphi) / d \Omega$
for the H2+H2/HD collisions as in \cite{guo02}
with the use of different PESs.


\vspace{0.5cm}
\noindent{{\bf Acknowledgment}}  
\vspace{2mm}
This work was supported by the St. Cloud State University (St. Cloud,
Minnesota, USA) internal grant program. 
%


\clearpage

\clearpage
\begin{list}{}{\leftmargin 2cm \labelwidth 1.5cm \labelsep 0.5cm}

\item[\bf Fig. 1] Elastic scattering cross section for
$ortho$-H$_2$+HD, $para$-H$_2$+HD and HD+HD at ultralow collsion energies. 
Calculations are done with the BMKP PES.

\item[\bf Fig. 2] Elastic scattering cross section for
HD+HD calculated with the use of the BMKP PES and the corresponding
result from work \cite{johnson79}.

\item[\bf Fig. 3] Rotational state resolved integral cross sections for
$ortho$-$\mbox{H}_2(j_2) +\mbox{HD}(j_1) \rightarrow \mbox{H}_2(j'_2)
  + \mbox{HD}(j'_1)$. 
Upper plot: initial states of HD and H$_2$ molecules are $j_1=j_2=1$
and corresponding final states are $j'_1=0$ and $j'_2=1$. 
In the bottom plots:
$j_1=2$, $j_2=1$ and the corresponding final states are $j'_1=j'_2=1$
and $j'_1=0$, $j'_2=1$. 
Calculations are done with the BMKP PES (bold lines), circles, diamonds and 
triangles up are corresponding results from work \cite{schaefer90}.

\item[\bf Fig. 4]Rotational state resolved thermal rate coefficients for
$ortho$-$\mbox{H}_2(j_2) +\mbox{HD}(j_1) \rightarrow \mbox{H}_2(j'_2)
  + \mbox{HD}(j'_1)$. In the upper plot the initial states 
of HD and H$_2$ molecules are $j_1=j_2=1$ 
and the corresponding final states are $j'_1=0, j'_2=1$. 
In the bottom plot, $j_1=2$, $j_2=1$ and corresponding final states are $j'_1=j'_2=1$. 
Calculations are done with the BMKP PES (bold lines), triangles up and
squares are corresponding results from work \cite{schaefer90}.
Results from work \cite{flower99a} are also shown.

\item[\bf Fig. 5] Rotational state resolved thermal rate coefficients for
$para$-$\mbox{H}_2(j_2) +\mbox{HD}(j_1) \rightarrow \mbox{H}_2(j'_2) +
  \mbox{HD}(j'_1)$. 
Upper plot: initial states of HD and H$_2$ molecules are $j_1=1$ and $j_2=0$ and 
corresponding final states are $j'_1=j'_2=0$. In the bottom plots:
$j_1=2$, $j_2=0$ 
and the corresponding final states are $j'_1=1$, $j'_2=0$ and
$j'_1=j'_2=0$. 
Calculations are done with the BMKP PES (bold lines), circles,
triangles up and diamonds 
are corresponding results from work \cite{schaefer90}.
Results from work \cite{flower99a} are also shown.
%

\end{list}

\clearpage
\begin{figure}
\begin{center}
\includegraphics*[scale=1.0,width=27pc,height=15pc]{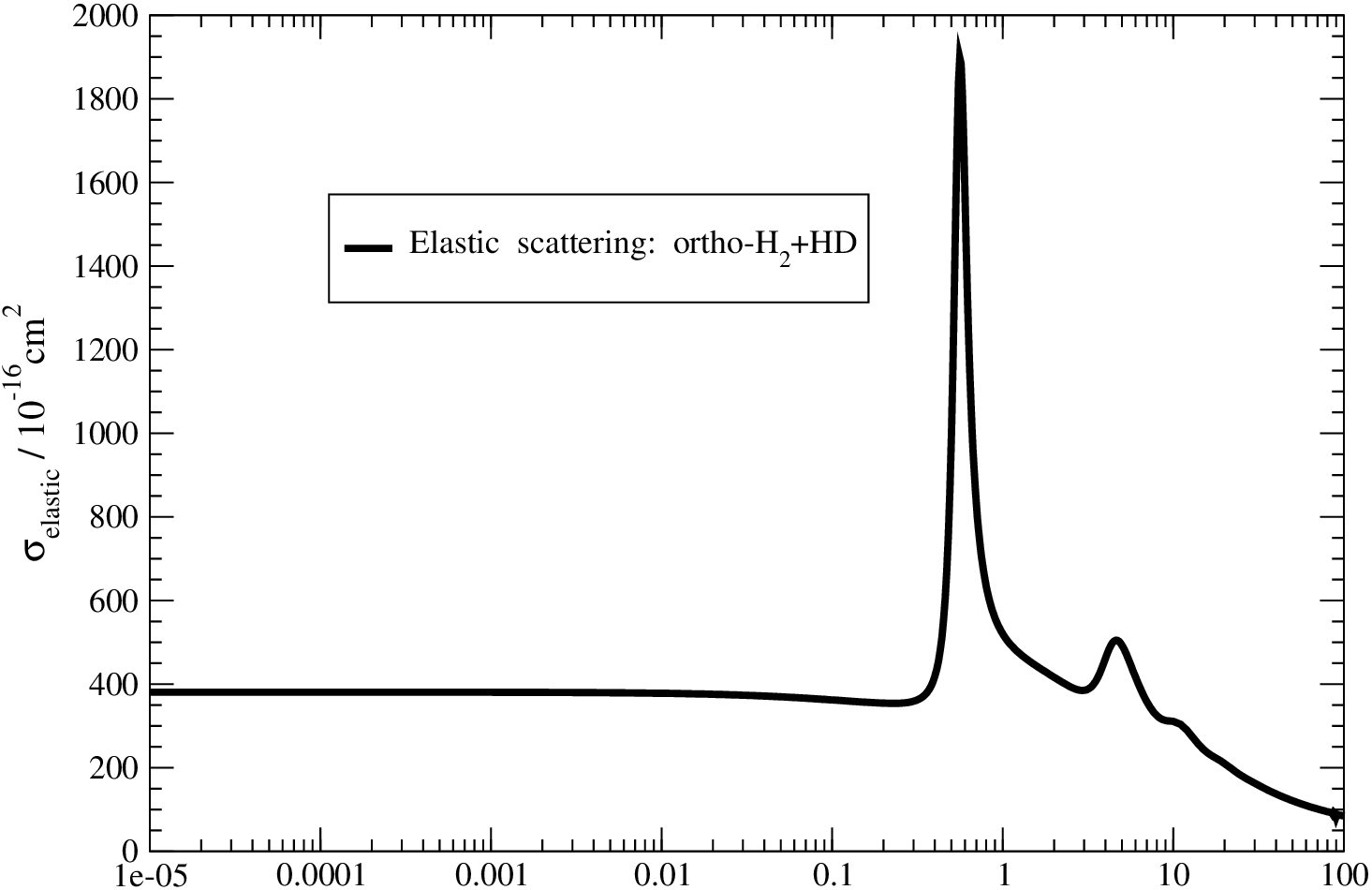}
\vspace{1mm}\\
\includegraphics*[scale=1.0,width=27pc,height=17pc]{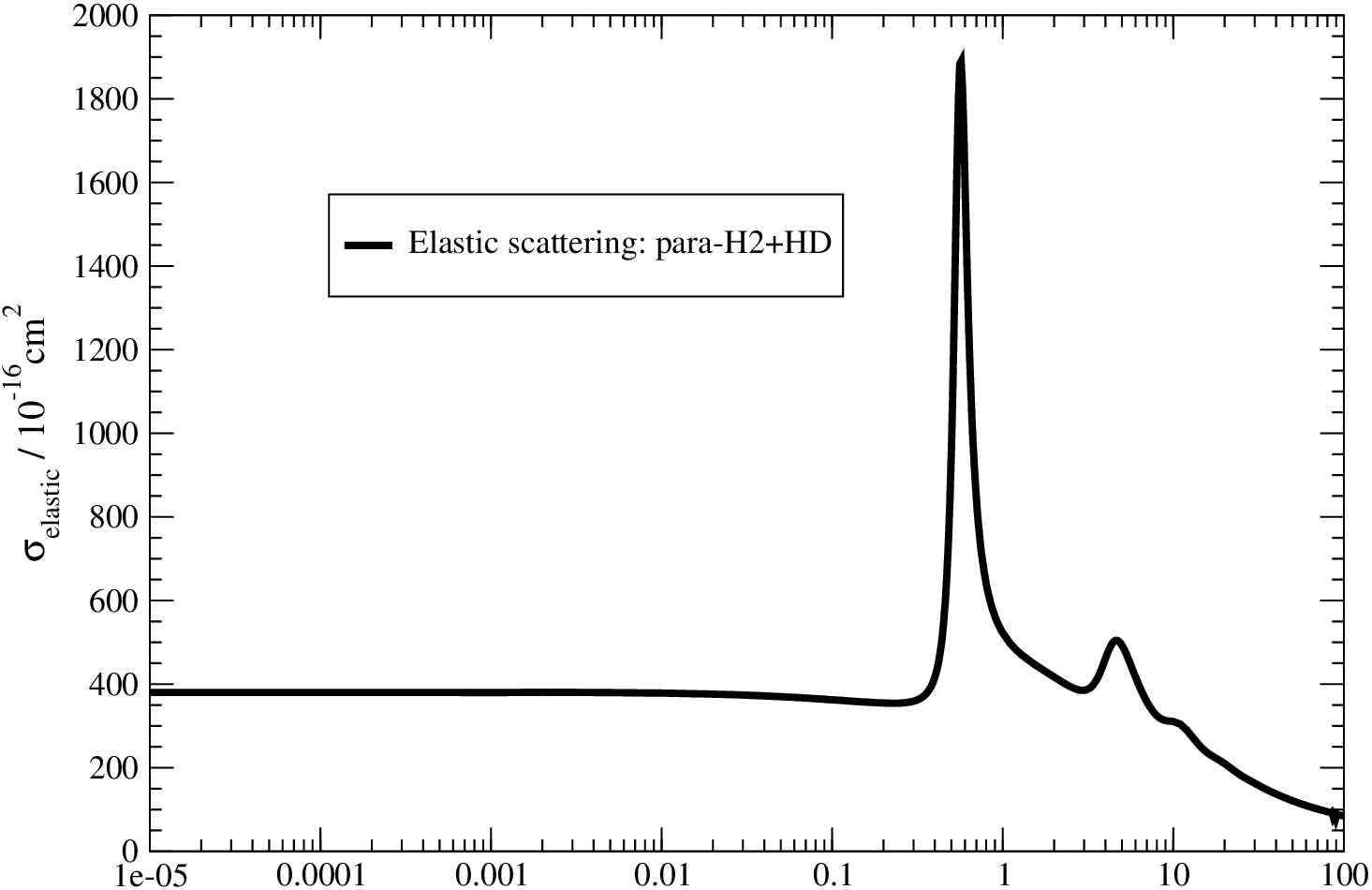}
\vspace{1mm}\\
\includegraphics*[scale=1.0,width=27pc,height=17pc]{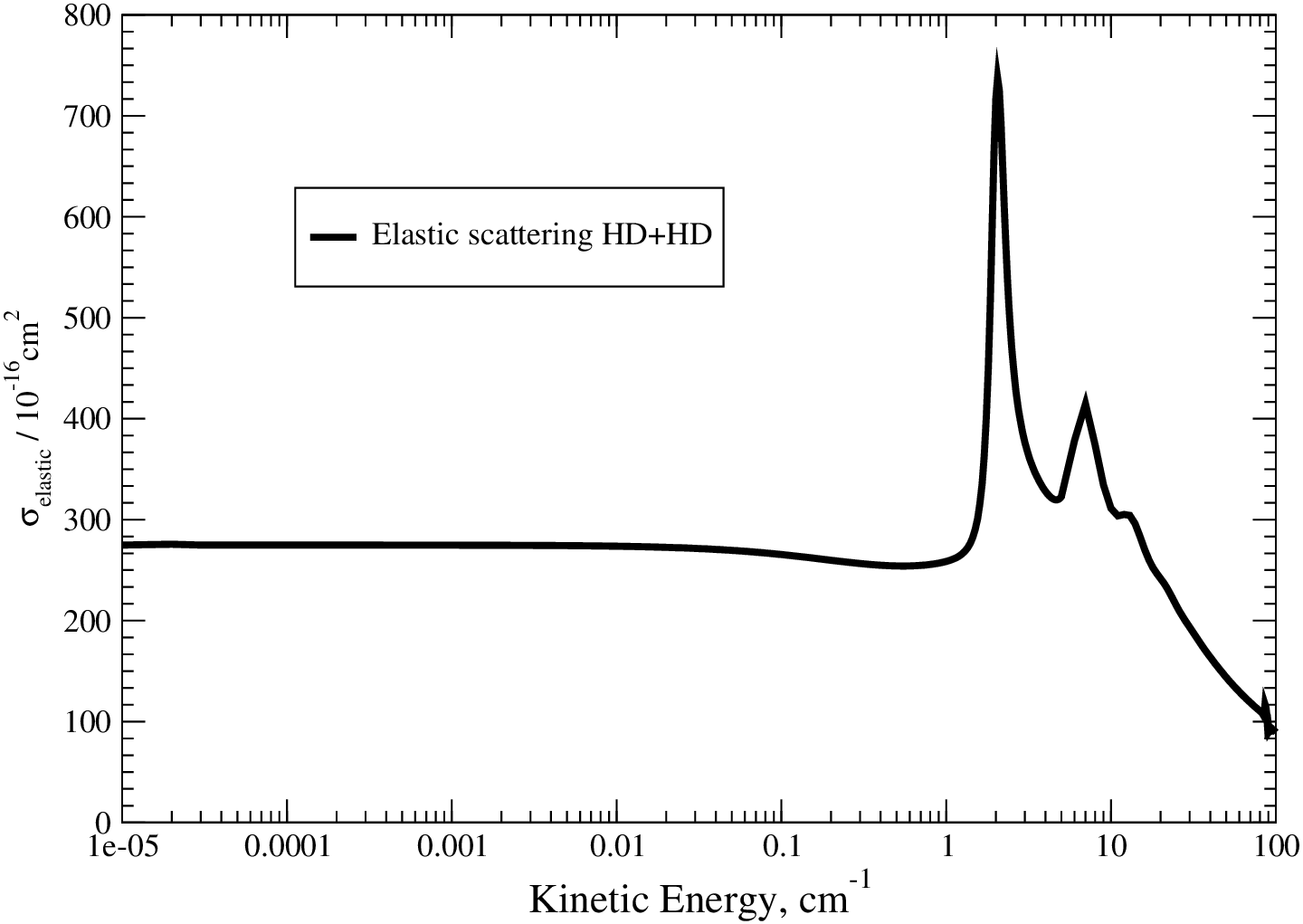}
\end{center}
%

\caption{Elastic scattering cross section for $ortho$-/$para$-H$_2$+HD 
and HD+HD at ultralow collsion energies. Calculations are done with the BMKP PES. 
}
\label{fig:fig1}
\end{figure}

\clearpage
\begin{figure}
\begin{center}
\includegraphics*[scale=1.0,width=27pc,height=15pc]{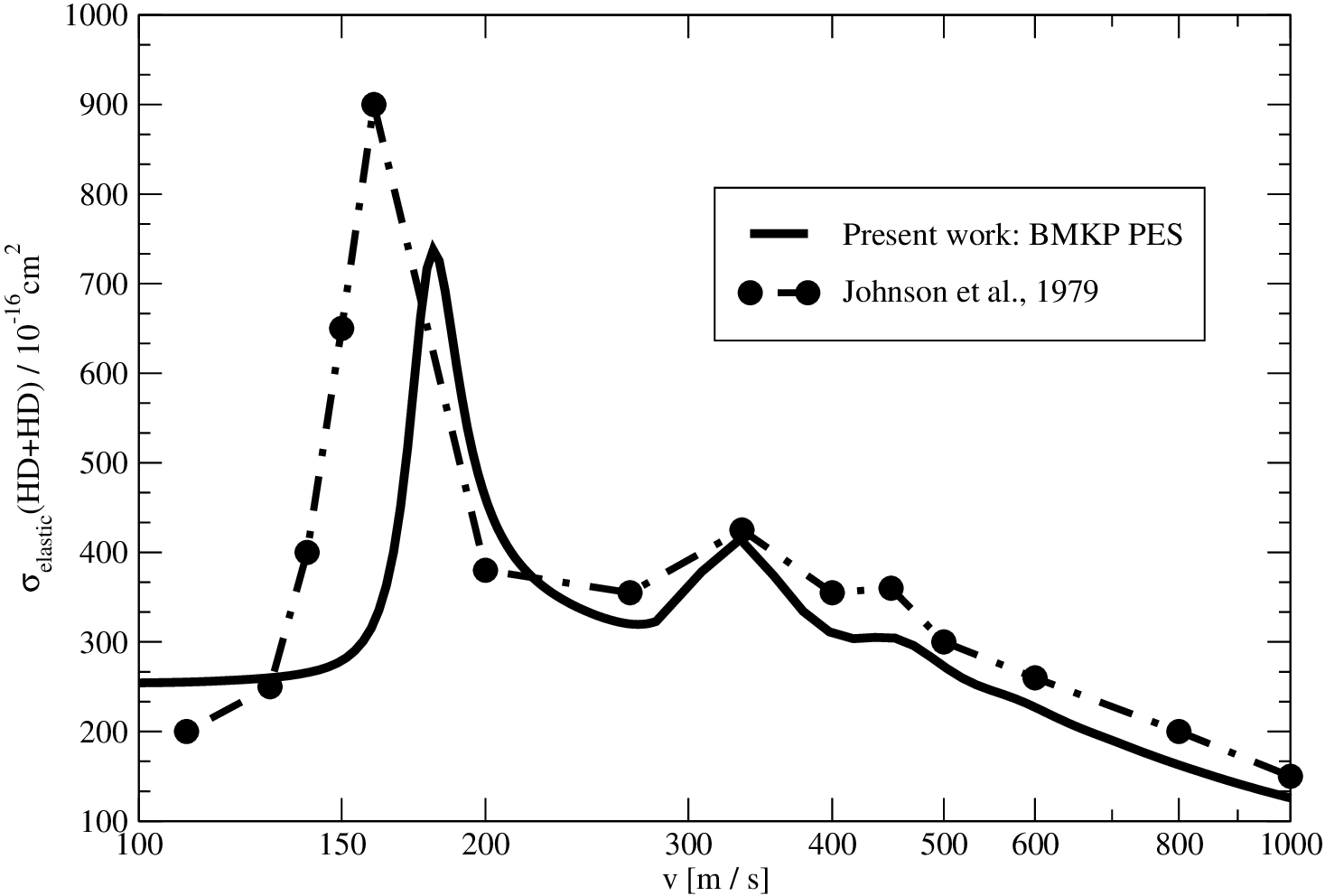}
\end{center}
\caption{Elastic scattering cross section for
HD+HD calculated with the use of the BMKP PES and the corresponding
result from work \cite{johnson79}.}
\label{fig:fig2}
\end{figure}

\clearpage
\begin{figure}
\begin{center}
\includegraphics*[scale=1.0,width=27pc,height=15pc]{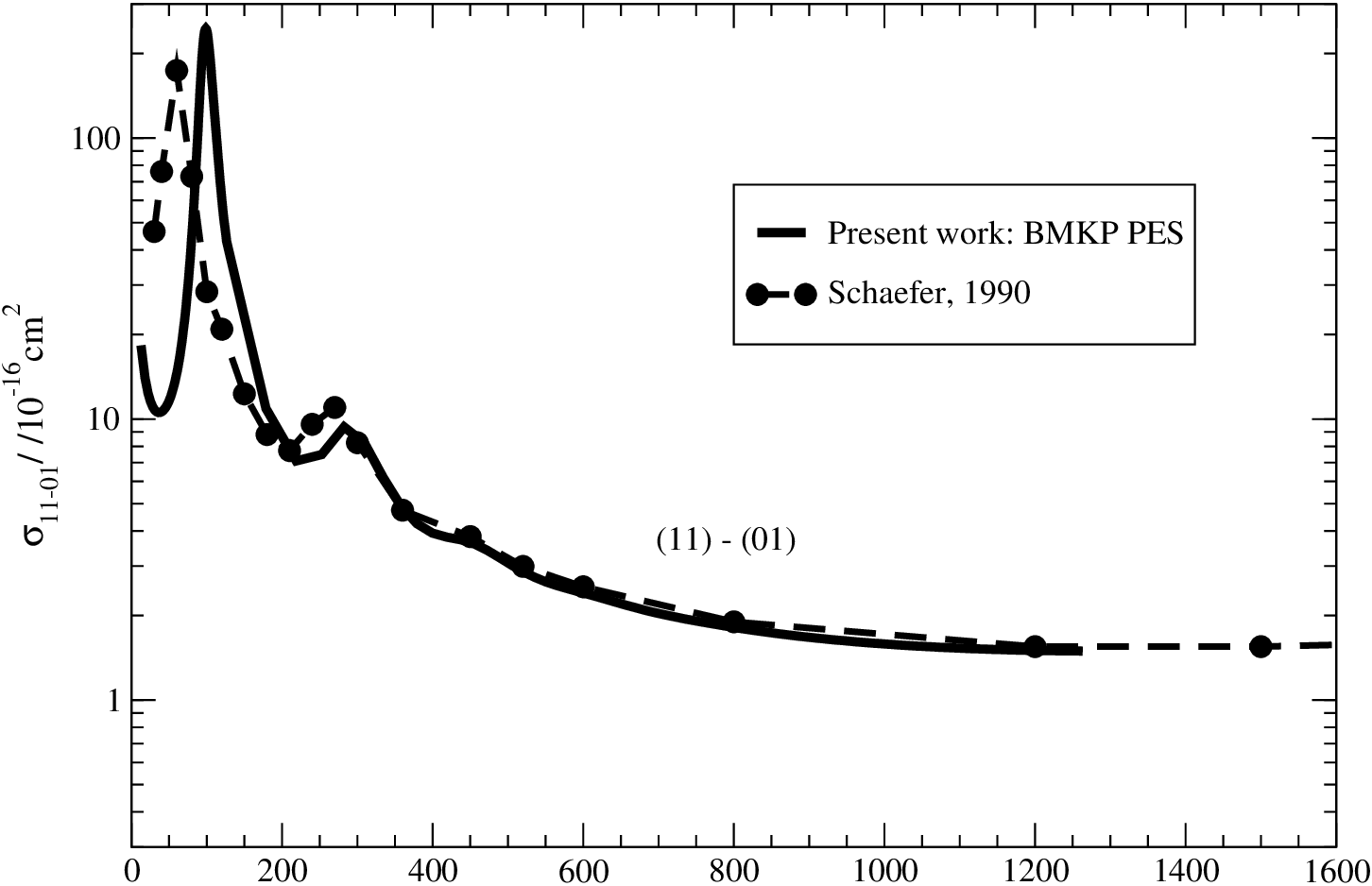}
\vspace{1mm}\\
\includegraphics*[scale=1.0,width=27pc,height=15pc]{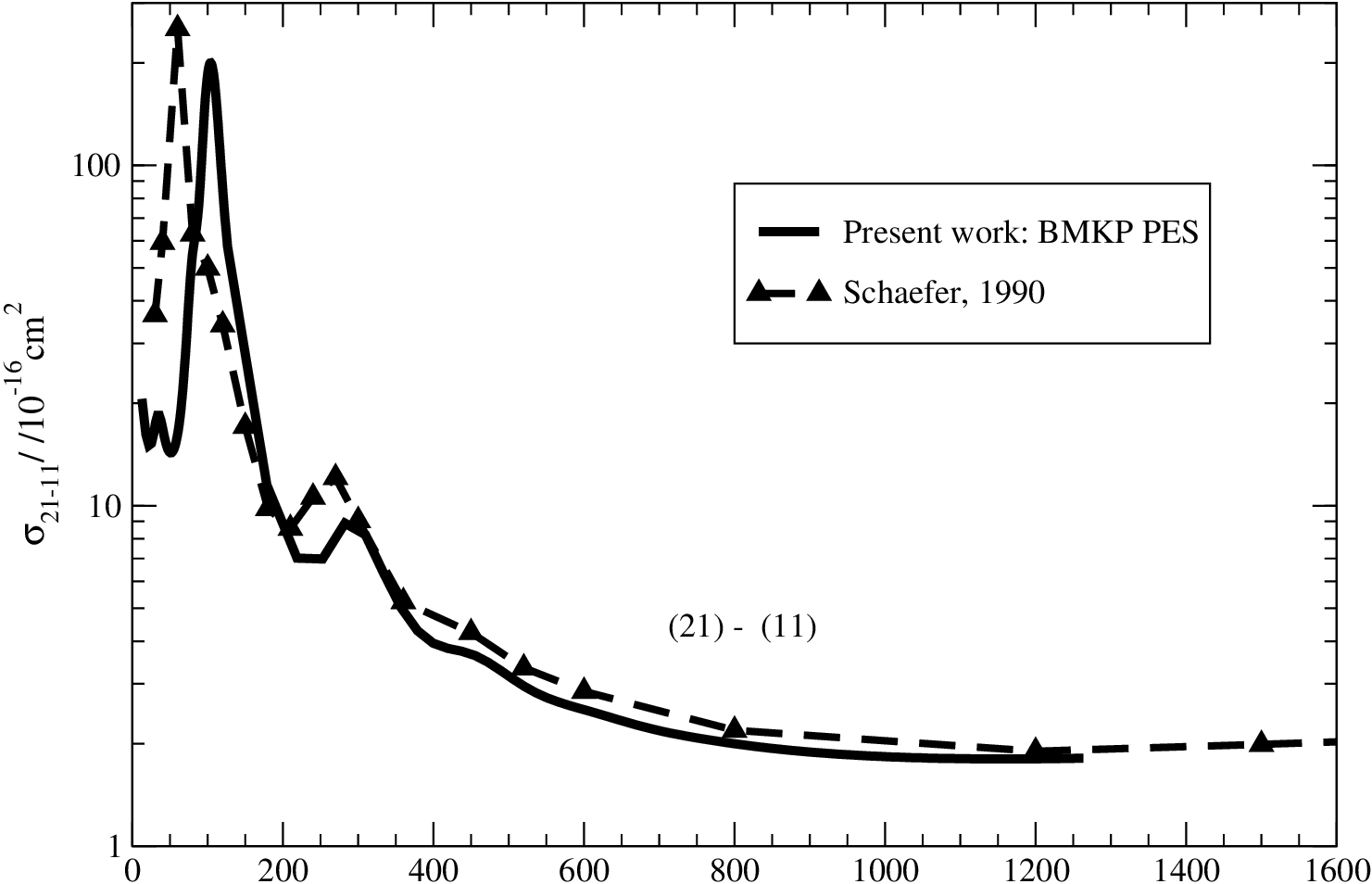}
\vspace{1mm}\\
\includegraphics*[scale=1.0,width=27pc,height=15pc]{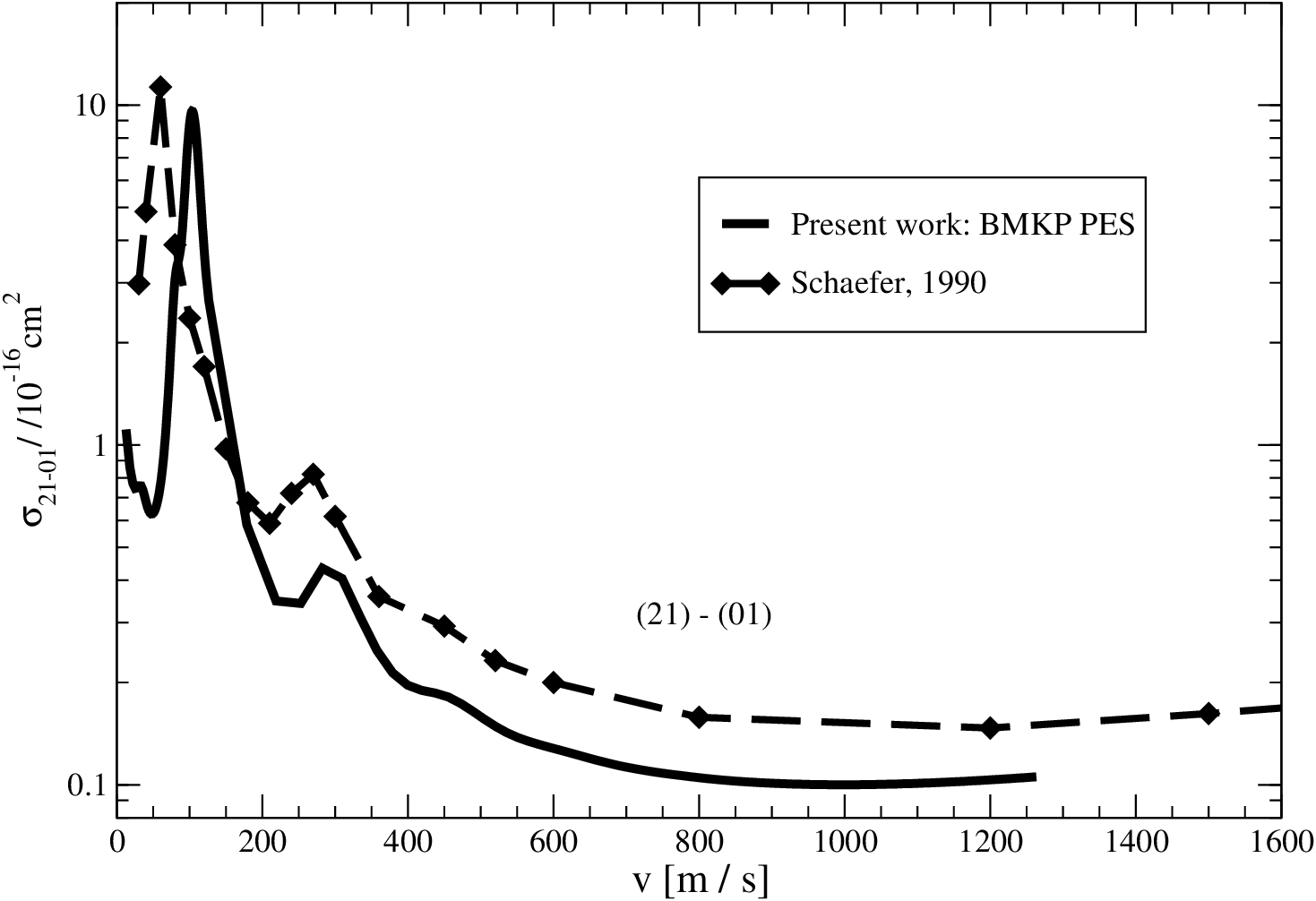}
\end{center}
\caption{Rotational state resolved integral cross sections for
$ortho$-$\mbox{H}_2(j_2) +\mbox{HD}(j_1) \rightarrow \mbox{H}_2(j'_2) + \mbox{HD}(j'_1)$.
Upper plot: initial states of HD and H$_2$ molecules are $j_1=j_2=1$ and 
corresponding final states are $j'_1=0$ and $j'_2=1$. In the bottom plots:
$j_1=2$, $j_2=1$ and the corresponding final states are $j'_1=j'_2=1$ and $j'_1=0$, $j'_2=1$.
Calculations are done with the BMKP PES (bold lines), circles, diamonds and triangles up are corresponding 
results from work \cite{schaefer90}.}
\label{fig:fig3}
\end{figure}

\clearpage
\begin{figure}
\begin{center}
\includegraphics*[scale=1.0,width=27pc,height=15pc]{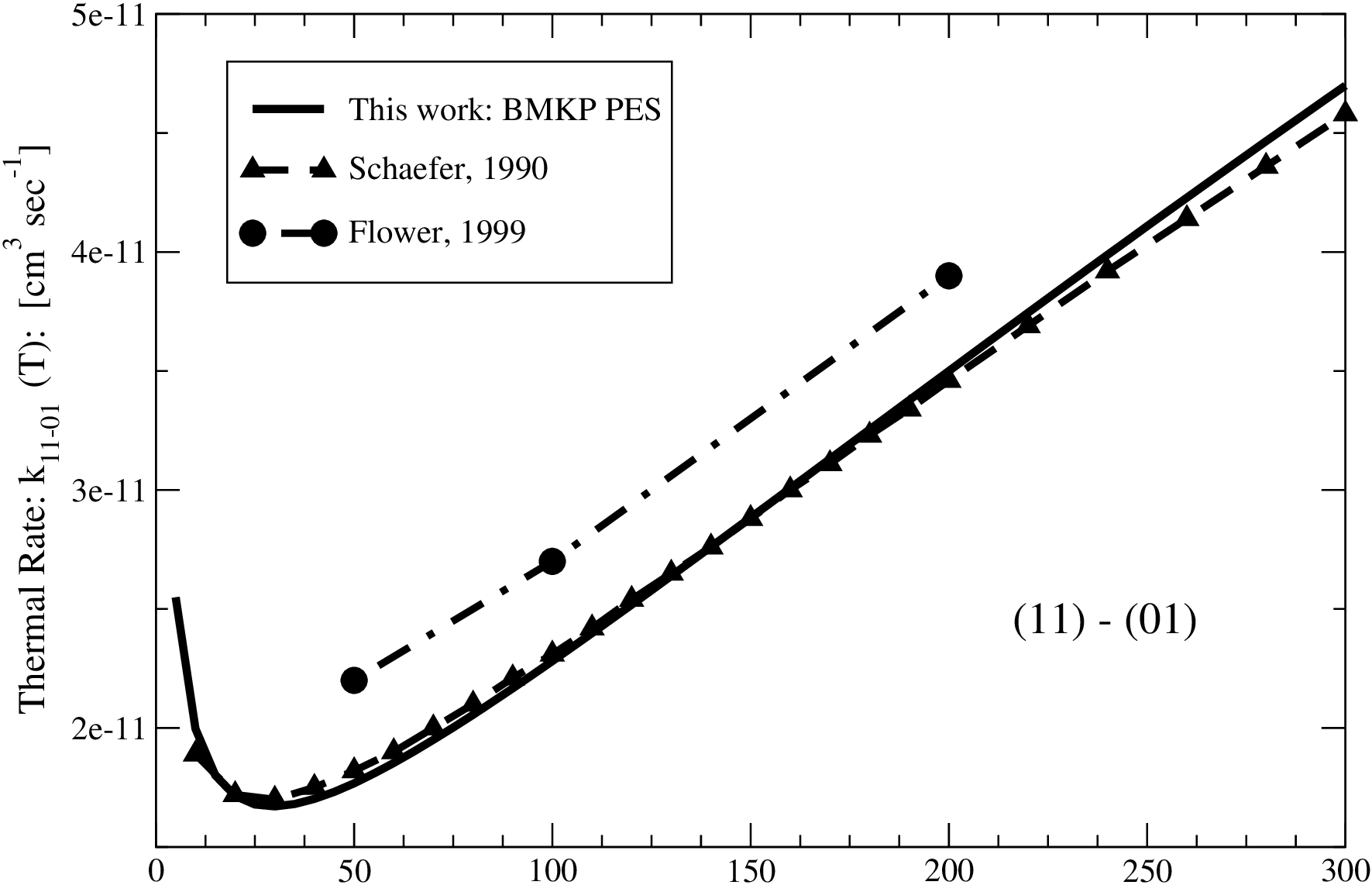}
\vspace{1mm}\\
\includegraphics*[scale=1.0,width=27pc,height=15pc]{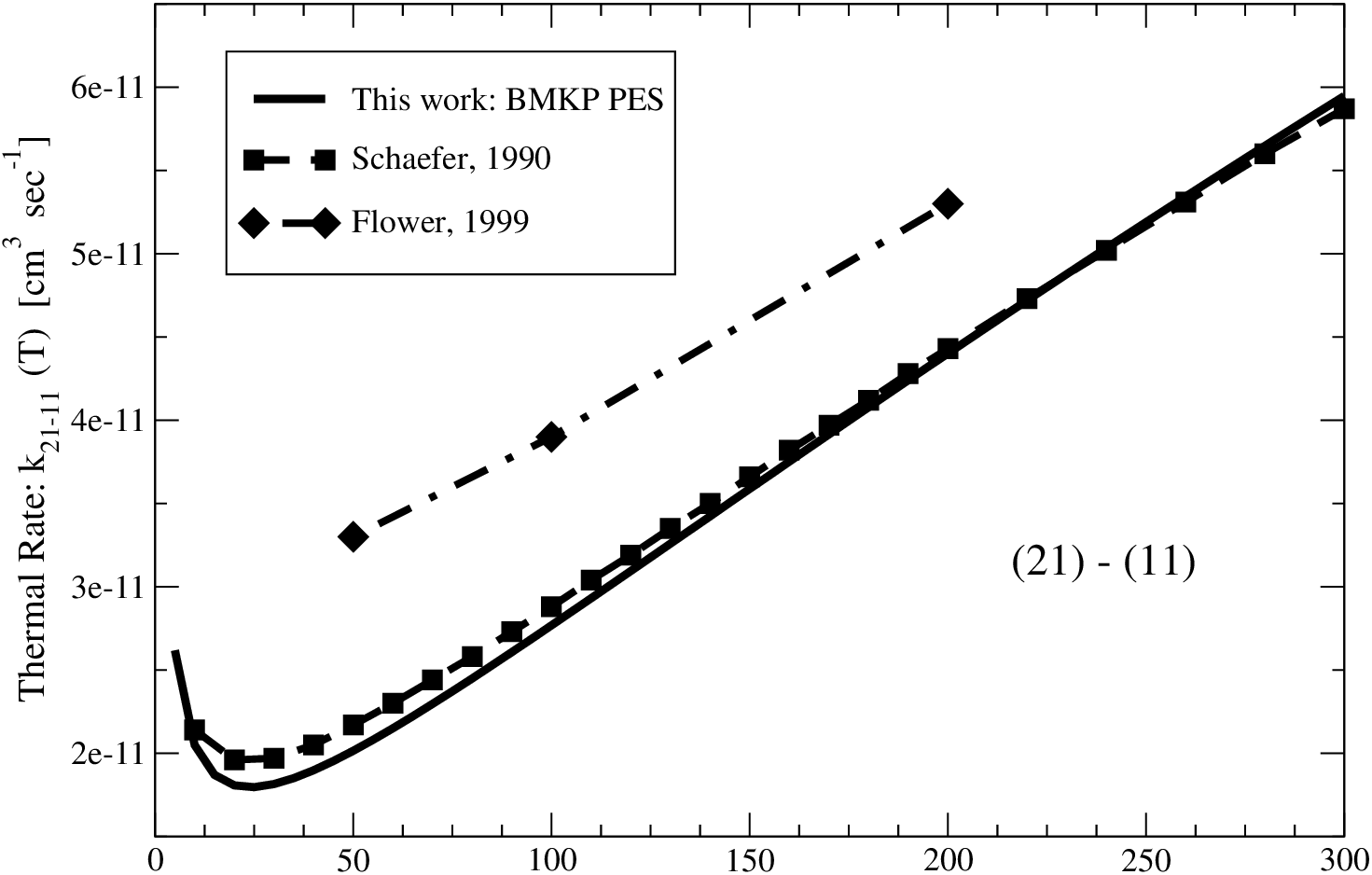}
\includegraphics*[scale=1.0,width=27pc,height=15pc]{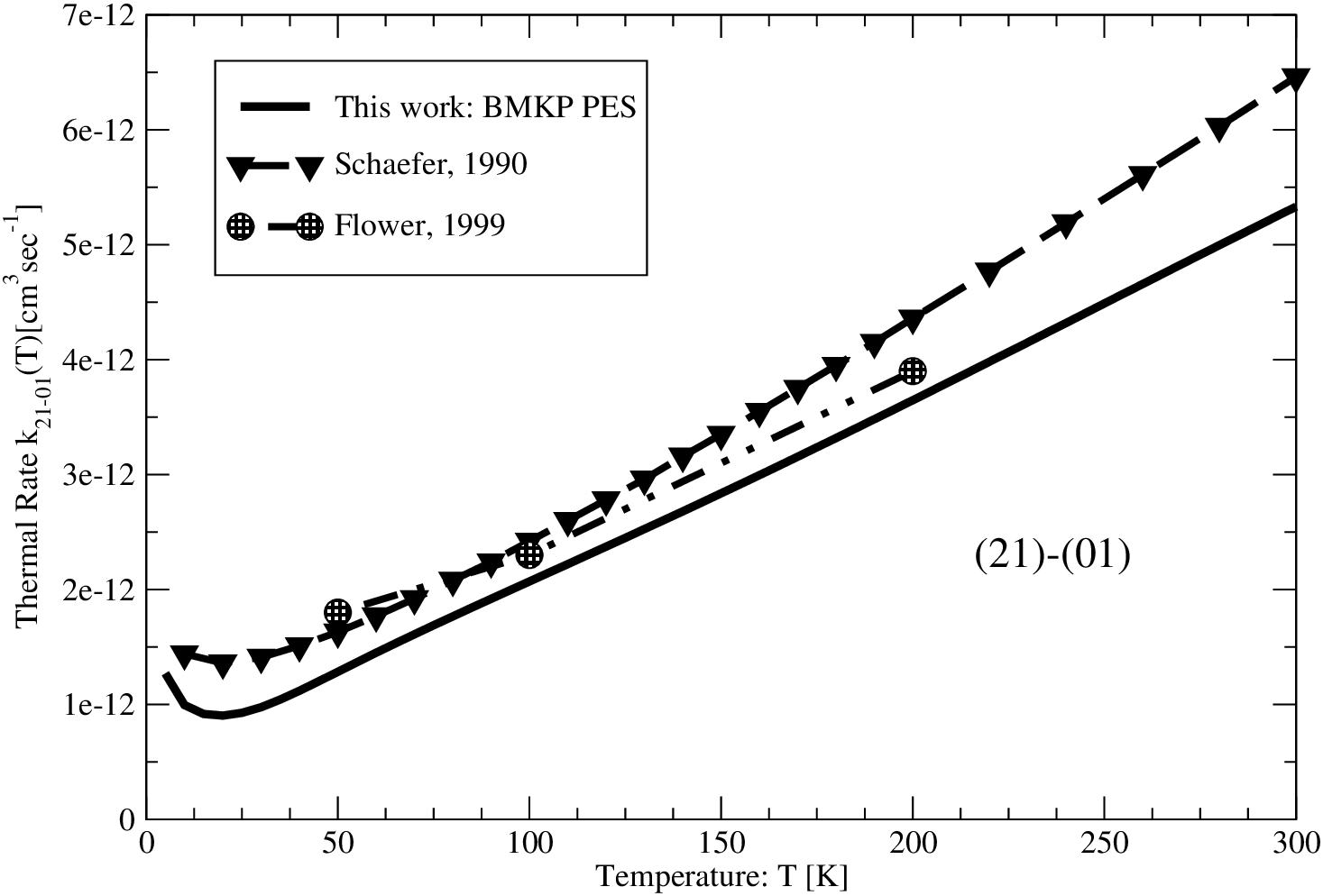}
\end{center}
\caption{Rotational state resolved thermal rate coefficients for
$ortho$-$\mbox{H}_2(j_2) +\mbox{HD}(j_1) \rightarrow \mbox{H}_2(j'_2) + \mbox{HD}(j'_1)$.
In the upper plot the initial states of HD and H$_2$ molecules are $j_1=j_2=1$ 
and the corresponding final states are $j'_1=0, j'_2=1$. In the bottom plot, $j_1=2$, $j_2=1$
and corresponding final states are $j'_1=j'_2=1$.
Calculations are done with the BMKP PES (bold lines), triangles up and squares are corresponding 
results from work \cite{schaefer90}.
Results from work \cite{flower99a} are also shown.}
\label{fig:fig4}
\end{figure}

\clearpage
\begin{figure}
\begin{center}
\includegraphics*[scale=1.0,width=27pc,height=15pc]{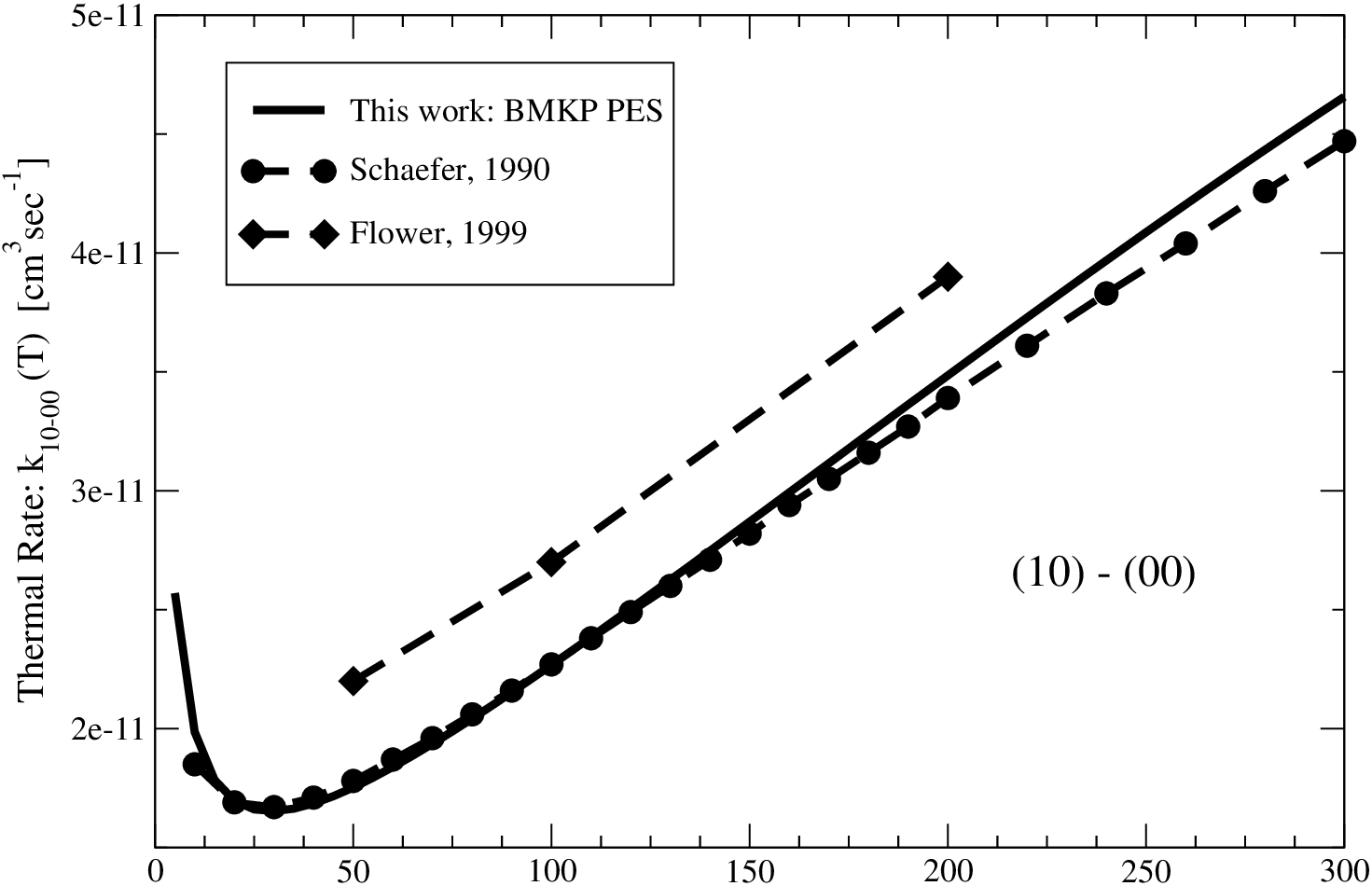}
\vspace{1mm}\\
\includegraphics*[scale=1.0,width=27pc,height=15pc]{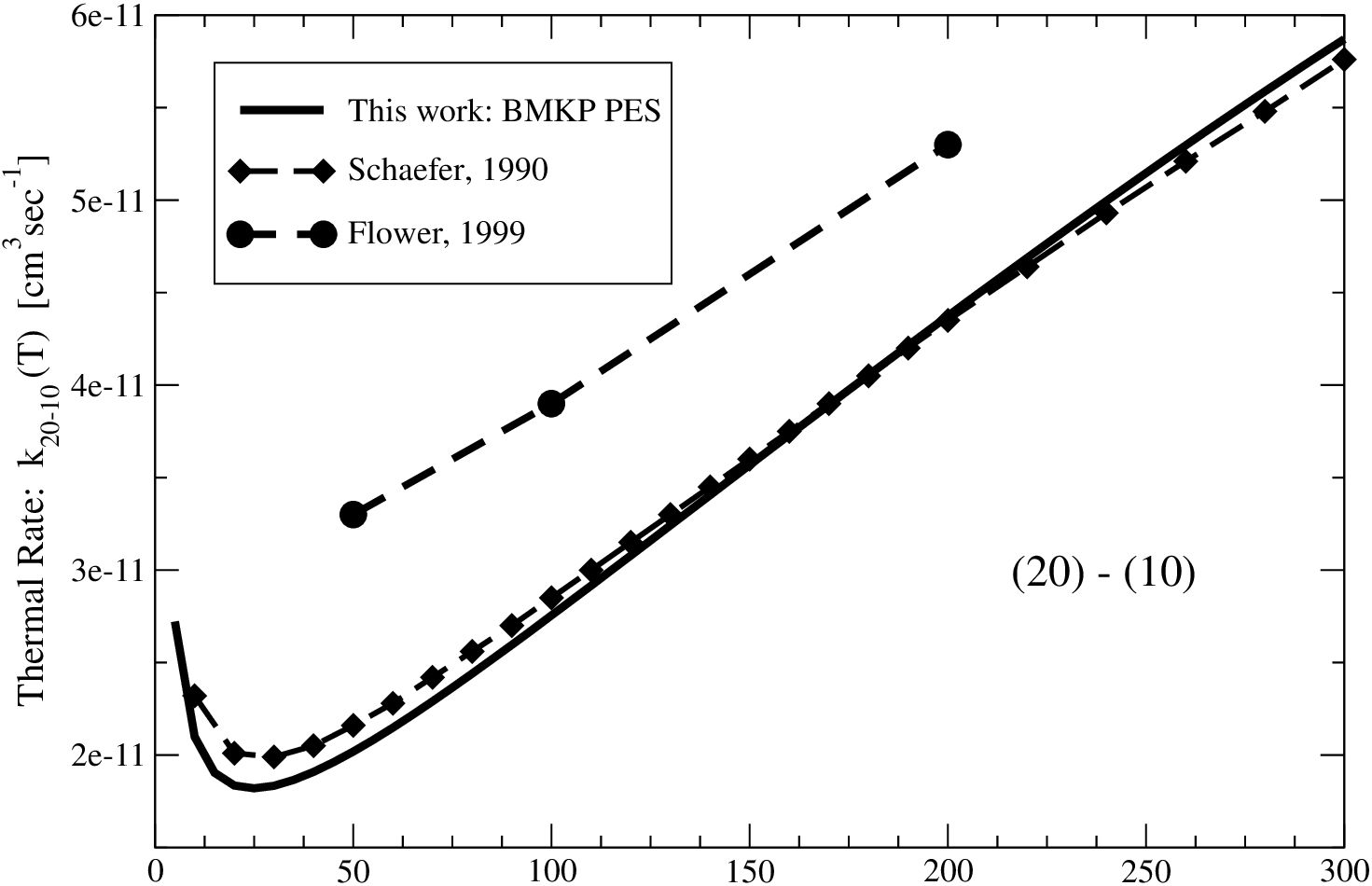}
\vspace{1mm}\\
\includegraphics*[scale=1.0,width=27pc,height=15pc]{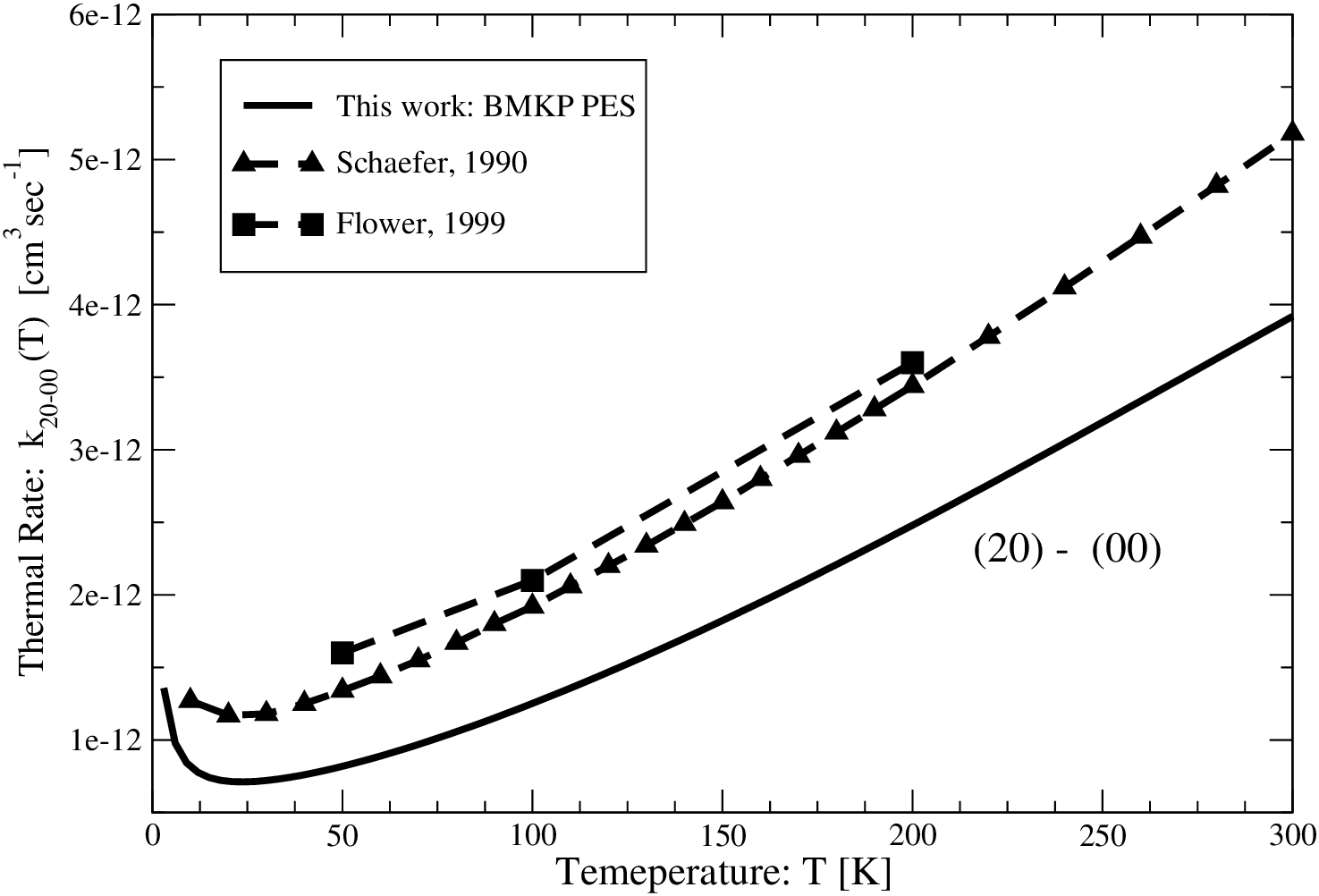}
\end{center}
\caption{Rotational state resolved thermal rate coefficients for
$para$-$\mbox{H}_2(j_2) +\mbox{HD}(j_1) \rightarrow \mbox{H}_2(j'_2) + \mbox{HD}(j'_1)$.
Upper plot: initial states of HD and H$_2$ molecules are $j_1=1$ and $j_2=0$ and 
corresponding final states are $j'_1=j'_2=0$. In the bottom plots:
$j_1=2$, $j_2=0$ and the corresponding final states are 
$j'_1=1$, $j'_2=0$ and $j'_1=j'_2=0$.
Calculations are done with the BMKP PES (bold lines), circles, triangles up and
diamonds are corresponding results from work \cite{schaefer90}.
Results from work \cite{flower99a} are also shown.}
%
\label{fig:fig5}
\end{figure}

\end{document}